\documentclass[12pt]{iopart}

\usepackage{graphicx}
\usepackage{dsfont}
\usepackage{iopams}
\usepackage{dsfont}
\usepackage{epstopdf}

\newcommand{\ket}[1]{\left|#1\right>}
\newcommand{\bra}[1]{\left<#1\right|}

\begin{document}

\title{Conditional generation of optical Schr\"odinger cat states}

\author{P. Adam$^{(1)}$, T. Kiss$^{(1)}$, Z. Dar\'azs$^{(1)}$ and I. Jex$^{(2)}$}
\address{$^{(1)}$ Department of Quantum Optics and Quantum Information, Research
Institute for Solid State Physics and Optics, Hungarian Academy of
Sciences, Konkoly-Thege M. u. 29-33, H-1121 Budapest, Hungary}
\address{$^{(2)}$ Department of Physics, FJFI \v CVUT v Praze, B\v
rehov\'a 7, 115 19 Praha 1 - Star\'e M\v{e}sto, Czech Republic}


\ead{adam@szfki.hu}

\begin{abstract}
Given a source of two coherent state superpositions with small separation in a traveling wave optical setting, we show that by interference and balanced homodyne measurement it is possible to conditionally prepare a symmetrically placed superposition of coherent states around the origo of the phase space. The separation of the coherent states in the superposition will be amplified during the process.   
\end{abstract}

\section{Introduction}
Preparation of superpositions of coherent states is a nontrivial task in traveling wave optics \cite{Glancy}.
There has been quite much progress in recent years: quantum states approximating small separation superpositions have been reported in experiments. These states would have important applications for quantum communication and information processing. Especially, coherent state superpositions placed symmetrically with respect to the origin (e.g. on the negative and positive part of the real axis) of the phase space (CSS)  are of interest.

One way to prepare CSSs is to apply photon subtraction from a squeezed vacuum state \cite{Dakna}. This method works well for approximating small separation CSSs as demonstrated in experiments \cite{CSSexperiments}. Extending this technique squeezed CSSs were generated with higher separation \cite{CSSexperiments2}. In another proposal for conditional preparation, squeezed single photon states are let interfere on a beamsplitter and states in one arm are conditionally selected by photon detection \cite{Lund2004}.

Nonlinearity due to the cross-Kerr effect has been proposed in an alternative scheme to produce 
superpositions of coherent states in a scheme by Gerry \cite{Gerry1999} where a one photon auxiliary state is utilized. The weakness of the Kerr effect allows for only a small separation of the states in the phase space. Moreover, these superpositions are not placed symmetric with respect to the origin, but have a large coherent amplitude. Separation of the constituent states can be increased at a cost of increasing the intensity of the input coherent beam \cite{Jeong2005}. One possibility to prepare CSS in such a scheme is to displace the states \cite{Jeong2005} by a high transmission beam splitter and a second coherent beam. Experimentally the intensities needed for the displacement do not appear realistic. Recently, a double cross-Kerr scheme have been proposed \cite{Bergou2009} where the auxiliary one photon state is in a superposition of two orthogonal polarizations. Weak self-Kerr effect was utilized in a conditional scheme with homodyne measurement in another proposal for CSS preparation \cite{JeongKim2004}. We note that the feasibility of schemes applying the Kerr nonlinearity are discussed in the literature, since a multimode, continuous-time model with causal, non-instantaneous response function establish limits on single photon operations \cite{Shapiro}.

In this paper we show that two independently prepared but identical superpositions of coherent states, displaced from the vacuum with a large amplitude, as prepared e.g. by Gerry's scheme, can be used prepare a CSS with increased separation. We apply the interference of the two states and a   conditional balanced homodyne measurement on one of the outputs.

\section{The scheme}
Let us assume that a source of coherent state superpositions of the following form are available
\begin{equation}
|\Phi_{0}\rangle={\cal N}_\Phi (|\alpha\rangle+|\alpha e^{-i\varphi}\rangle)\, 
\end{equation}
where the separation of the superposition is characterized by $d_0=|\alpha-\alpha e^{-i\varphi}|=2\alpha_0 \sin \varphi/2\approx \alpha_0 \varphi$ for small $\varphi$, with the notation $\alpha_0=|\alpha|$. The normalization constant is denoted by ${\cal N}_\Phi$, throughout the paper similar notation will be applied.
Our aim is to shift the superposition onto the real axis, symmetrically arranged around the origin and at the same time to increase the separation of the two constituent coherent states.
\begin{figure}[ht!]
\label{fig:scheme}
\center
\resizebox{0.77 \linewidth}{!}{
 \includegraphics{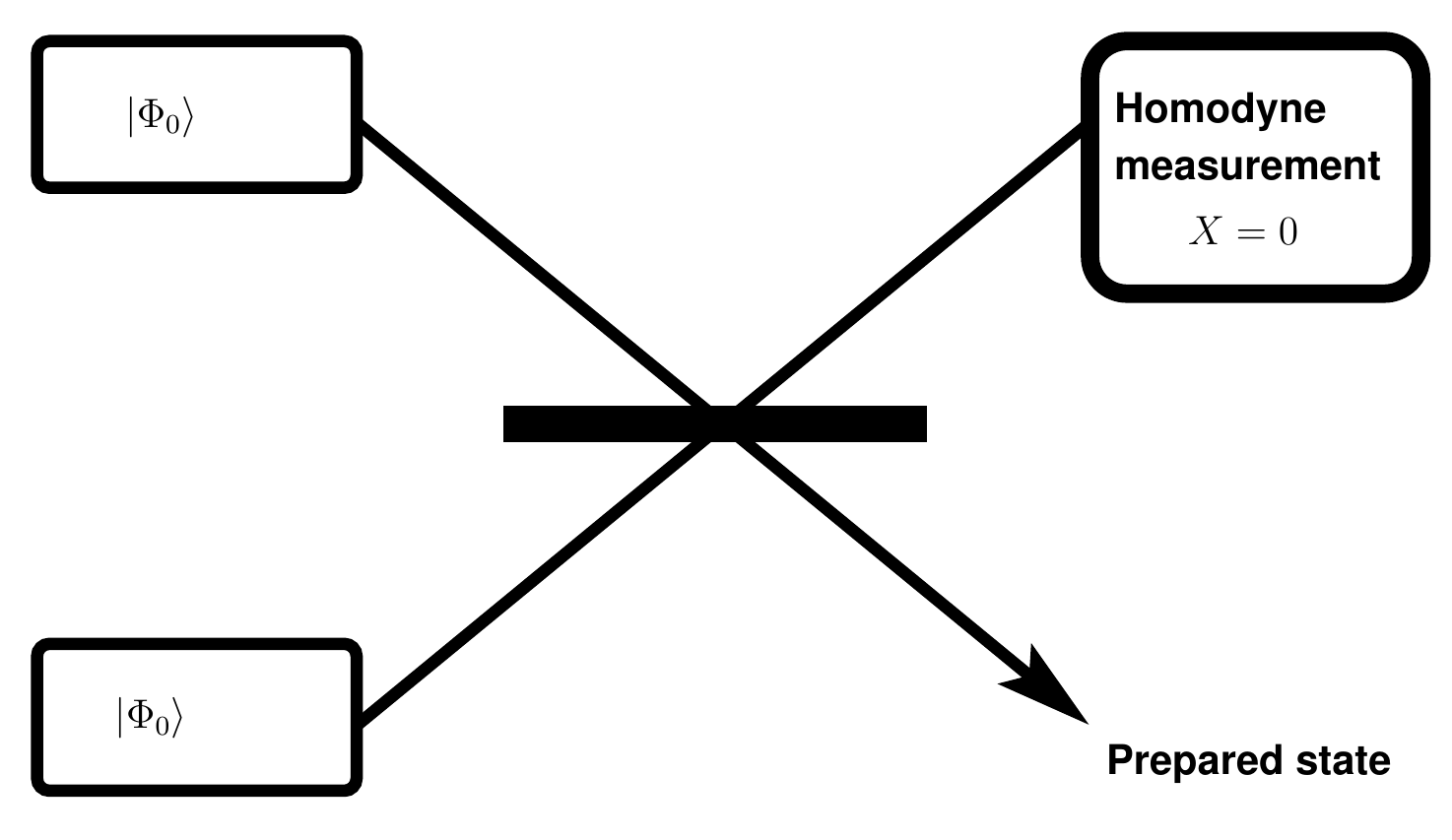}}
 \caption{The proposed scheme: two identical superpositions of coherent states are let to interfere on a balanced beam splitter, then the $X$ quadrature is measured on one of them and the $0$ result selects the prepared state.}
\end{figure}
Let us consider two identically prepared initial states falling on a symmetric beam splitter as shown in Figure 1.
The beam splitter transformation is defined by its action on coherent states
\begin{equation}
\ket{\alpha}_{1}\ket{\beta}_{2} \to \ket{\frac{1}{\sqrt{2}}(\alpha+\beta)}_{3} \ket{\frac{1}{\sqrt{2}}(\alpha-\beta)}_{4} \, .
\end{equation}
For simplicity, we can fix the phase of  the initial coherent state superposition $\arg \alpha = \pi/2+\varphi/2$, without losing generality.
The transformed outgoing two mode state then reads 
\begin{eqnarray}
\label{eq:psi34}
\ket{\Psi}_{34}&=&
{\cal N}_{\Psi}\left[\left(
\ket{\sqrt{2} i \alpha_0 e^{i \varphi/2}}_{3}+
\ket{\sqrt{2} i \alpha_0 e^{-i \varphi/2}}_{3}
\right) \ket{0}_{4} + 
\right. \nonumber\\ &+& \left.
\frac{1}{\sqrt{2}}\ket{\sqrt{2} i \alpha_0 \cos \frac{\varphi}{2}}_{3}
\ket{cat}_{4}\right]\, .
\end{eqnarray}
The following Schr\"odinger cat state is the desired superposition on the real axis of the phase space
\begin{eqnarray}
\ket{cat}_4&=&{\cal N}_{cat}\left( 
\ket{\sqrt{2}\alpha_0 \sin \frac{\varphi}{2}}_{4}
+ 
\ket{-\sqrt{2}\alpha_0 \sin \frac{\varphi}{2}}_{4}
\right)\, .
\end{eqnarray}
The above state is a CSS, with an increased separation $d=\sqrt{2}\alpha_0 \varphi\approx \sqrt{2} d_0$.
The next step of the preparation is the selection of the cat state in mode 4 by an appropriate measurement in mode 3. We show that balanced homodyne measurement of the $X$ quadrature (the real axis of the phase space) can provide a perfect selection, depending on the parameters of the initial states.

\section{Selection by balanced homodyne measurement}

Ideal, balanced homodyne detection measures the quadrature $X$, the phase of which being set by the auxiliary strong, classical laser beam. Let us consider the measurement of the real quadrature with the measurement result $X=0$. Since we have already fixed the phase of the initial state to be $\arg \alpha = \pi/2+\varphi/2$, the phase of the measured quadrature can be fixed as well. The projection corresponding to the measurement $|X=0\rangle \langle X=0| dX$ in mode 3 will select the second term form the superposition (\ref{eq:psi34})  with high probability  according to the symmetrical Gauss distribution around the origin. On the other hand, the contribution of the first term can be minimal, in fact zero.

The state in mode 4 after the measurement will be in the following state
\begin{equation}
\label{eq:prepared}
\ket{\Psi}_{4}={\mathcal N}(c_{1} \ket{0}_{4}+c_{2} \ket{cat}_{4})\, .
\end{equation}
The coefficients of the superposition are determined by the overlaps
\begin{eqnarray}
c_{1}&=&\bra{X=0}\alpha^{\prime}\rangle+\bra{X=0}\alpha^{\prime}e^{-i \varphi}\rangle dX
\\
c_{2}&=&\bra{X=0}\alpha^{\prime}(e^{-i \varphi}+1)/2\rangle dX\, ,
\end{eqnarray}
and the norm is $\mathcal N=(|c_{1}|^{2}+|c_{2}|^{2}+c_1^*c_2 \langle 0|cat\rangle+c_1 c_2^* \langle cat| 0 \rangle)^{-1}$. We can explicitly calculate the coefficients and look for the cases when $c_1$ is zero or can be neglected with respect to $c_2$. Since the second coefficient is a constant, $c_2=\pi^{-1/4}$, the functional form of $c_1$ will determine their ratio
\begin{equation}
\frac{|c_1|}{|c_2|}=\left| \exp\left[\left(-1+e^{-i \varphi }\right) \alpha_0 ^2 \right]+\exp\left[\left(-1+e^{i \varphi }\right) \alpha_0 ^2 \right] \right| \, .
\end{equation}
For small values of $\varphi$, which are of interest for us, one can expand the exponent in the exponential. In first order, this yields the following simple cosine function
\begin{equation}
\frac{|c_1|}{|c_2|}\approx 2 \left|\cos\left(\alpha_0 ^2 \varphi \right)\right| \, .
\end{equation}
For a given separation angle $\varphi$ in the original superposition, there will be a minimal optimum coherent amplitude $\alpha_0$ for which $c_1$ will become exactly zero as shown in Figure 2. 
\begin{figure}[ht!]
\label{fig:abra}
\center
\resizebox{0.7 \linewidth}{!}{
 \includegraphics{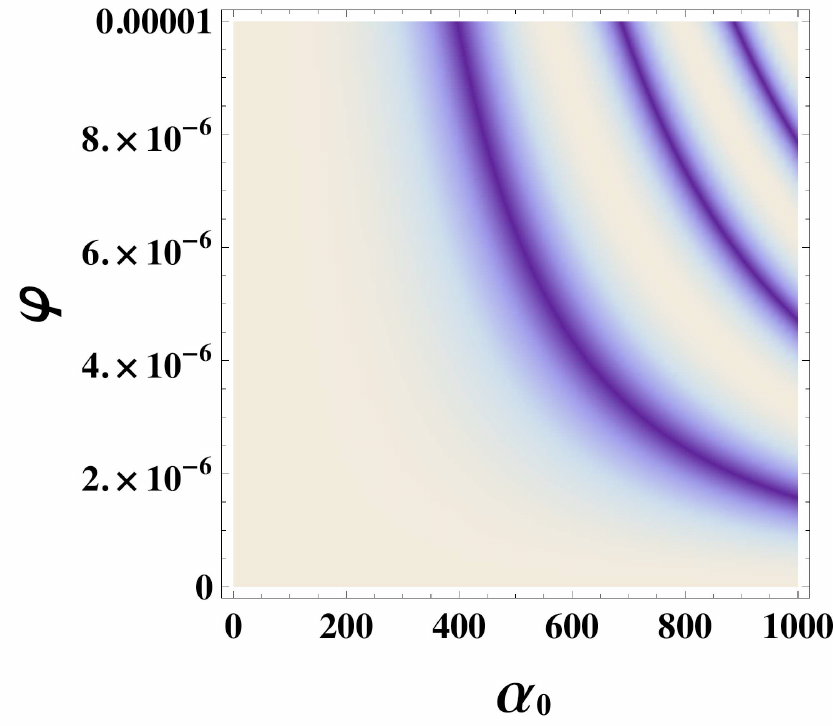}}
 \caption{The ratio of the coefficients $|c_1/c_2|$ in the prepared state (\ref{eq:prepared}) as a function of the initial state parameters $\alpha_0$ and $\varphi$. Dark color indicates values close to zero.}
\end{figure}
In first order approximation the first optimum reads
\begin{equation}
\alpha_0^{min}\approx \sqrt{\frac{\pi}{2 \varphi}} \, .
\end{equation}
For realistic values of $\varphi\lesssim 10^{-5}$ this leads to rather small separations $d\approx \sqrt{\pi \varphi}$.
The next order in the expansion will not change the zeros of the function, but introduces an exponential decay of the oscillation. The second order approximation reads
\begin{equation}
\frac{|c_1|}{|c_2|}\approx  e^{-\frac{1}{2} \alpha_0^2 \varphi ^2} 2 \left|\cos\left(\alpha_0 ^2 \varphi \right)\right|= e^{-\frac{d^2}{4}} 2 \left|\cos\left(\alpha_0 d \right)\right|\, .
\end{equation}
The last expression shows that for desirably large separations ($d\gtrsim4$) the ratio will be close to zero, thus the first term in (\ref{eq:prepared}) can be safely neglected.

\section{Conclusions}
In summary, we have proposed to apply balanced homodyne measurement to conditionally prepare traveling wave CSS states from a source of general coherent state superpositions. The method utilizes the destructive interference on a beam splitter to displace the superposition around the origin of the phase space, while the separation of the constituent coherent states is increased. the method could be used as a second step in any scheme where small separation coherent state superpositions are prepared with high coherent amplitude, as in \cite{Gerry1999}.  

The ideally sharp homodyne measurement of the quadrature assumed in the present paper would formally lead to zero probability of the preparation. Our calculations show \cite{inprep} that taking into account a finite detection window, feasible probabilities can be achieved while keeping the fidelity of the preparation high.

\ack
T. K. thanks A. Miranowicz for interesting discussions.
The financial support by MSM 6840770039, M\v SMT LC 06002, the Czech-Hungarian cooperation project (KONTAKT,CZ-10/2007) and by the Hungarian Scientific Research Fund (Contract No. T049234) is gratefully acknowledged.

\end{document}